\documentclass{article}

\usepackage[preprint,nonatbib]{nips_2018}

\usepackage[utf8]{inputenc} 
\usepackage[T1]{fontenc}    
\usepackage{hyperref}       
\usepackage{url}            
\usepackage{booktabs}       
\usepackage{amsfonts}       
\usepackage{nicefrac}       
\usepackage{microtype}      

\usepackage[pdftex]{graphicx}
\usepackage[numbers]{natbib}

\usepackage{fixltx2e}
\newcommand{\rpm}{\raisebox{.2ex}{$\scriptstyle\pm$}}
\usepackage{gensymb}
\usepackage[subpreambles=true]{standalone}
\usepackage{import}
\graphicspath{ {./figures/} }

\title{ A neural network trained to predict future video frames mimics critical properties of biological neuronal responses and perception}

\author{
  William Lotter$^1$, Gabriel Kreiman$^1$, David Cox$^{1,2,3}$\\
  $^1$Harvard University, $^2$MIT-IBM Watson AI Lab, $^3$IBM Research \\
{\tt\small lotter.bill1@gmail.com, gabriel.kreiman@tch.harvard.edu, david.d.cox@ibm.com}
}

\begin{document}

\maketitle

\begin{abstract}
While deep neural networks take loose inspiration from neuroscience, it is an open question how seriously to take the analogies between artificial deep networks and biological neuronal systems.
Interestingly, recent work has shown that deep convolutional neural networks (CNNs) trained on large-scale image recognition tasks can serve as strikingly good models for predicting the responses of neurons in visual cortex to visual stimuli, suggesting that analogies between artificial and biological neural networks may be more than superficial.
However, while CNNs capture key properties of the average responses of cortical neurons, they fail to explain other properties of these neurons.
For one, CNNs typically require large quantities of labeled input data for training.
Our own brains, in contrast, rarely have access to this kind of supervision, so to the extent that representations are similar between CNNs and brains, this similarity must arise via different training paths. 
In addition, neurons in visual cortex produce complex time-varying responses even to static inputs, and they dynamically tune themselves to temporal regularities in the visual environment.
We argue that these differences are clues to fundamental differences between the computations performed in the brain and in deep networks.
To begin to close the gap, here we study the emergent properties of a previously-described recurrent generative network that is trained to predict future video frames in a self-supervised manner.
Remarkably, the model is able to capture a wide variety of seemingly disparate phenomena observed in visual cortex, ranging from single unit response dynamics to complex perceptual motion illusions.
These results suggest potentially deep connections between recurrent predictive neural network models and the brain, providing new leads that can enrich both fields.

\end{abstract}

\section{Introduction}

The fields of neuroscience and machine learning have long enjoyed productive dialogue, with neuroscience offering inspiration for how artificial systems can be constructed, and machine learning providing tools for modeling and understanding biological neural systems.
Recently, as deep convolutional neural networks (CNNs) have emerged as leading systems for visual recognition tasks, these same models have emerged---without any modification or tailoring to purpose---as leading models for explaining the population responses of neurons in primate visual cortex \cite{Yamins_2014, Yamins_2016, Kriegeskorte_2014}.
These results suggest that the connections between artificial deep networks and brains may be more than skin deep.

However, while deep CNNs capture some important details of the responses of visual cortical neurons, they fail to explain other key properties of the brain.
Notably, the level of strong supervision used to train state-of-the-art CNNs is much greater than that available to our brain.
To the extent that representations in the brain are similar to those in CNNs trained on e.g. ImageNet, the brain must be arriving at these representations by different, largely unsupervised routes.
Another key difference is that CNNs are fundamentally static and lack a notion of time, whereas neuronal systems are highly dynamic, producing responses that vary dramatically in time, even in response to static inputs.
Figure \ref{on_off}a shows a typical response profile of a visual cortical neuron to a static input \cite{Schmolesky_1998}.
The neuron produces a brief transient response to the onset of the visual stimulus, followed by near total suppression of that response.
When the stimulus is removed, the neuron responds again with a transient burst of activity (known as an ``off'' response).
Neurons throughout visual cortex show a variety of dynamic response profiles, and the computational purpose of these dynamics is currently not well understood.

To further complicate matters, the responses of neurons in the primate visual cortex are also sensitive to long range temporal structure in the visual world.
For instance, Meyer and Olson \cite{Meyer_2011} showed that neurons in inferior temporal cortex (IT) could be strongly modulated by prior experience with sequences of presented images.
After repeated presentations of arbitrary images with predictable transition statistics (e.g. ``image B always follows image A''), neurons appeared to learn the sequence statistics, responding robustly only to sequence transitions that were unexpected.
The importance of temporal context in perception is further illustrated in various motion illusions, such as the flash-lag effect \cite{Nijhawan_1994, Mackay_1958, Eagleman_2000} and static motion illusions \cite{Watanabe_2018}, where the motion of objects is incorrectly perceived by humans in predictable ways.
Again, standard feedforward CNNs are insufficient to explain these temporal phenomena.

\begin{figure}[h]
  \begin{center}
    \includegraphics[width=0.62\textwidth]{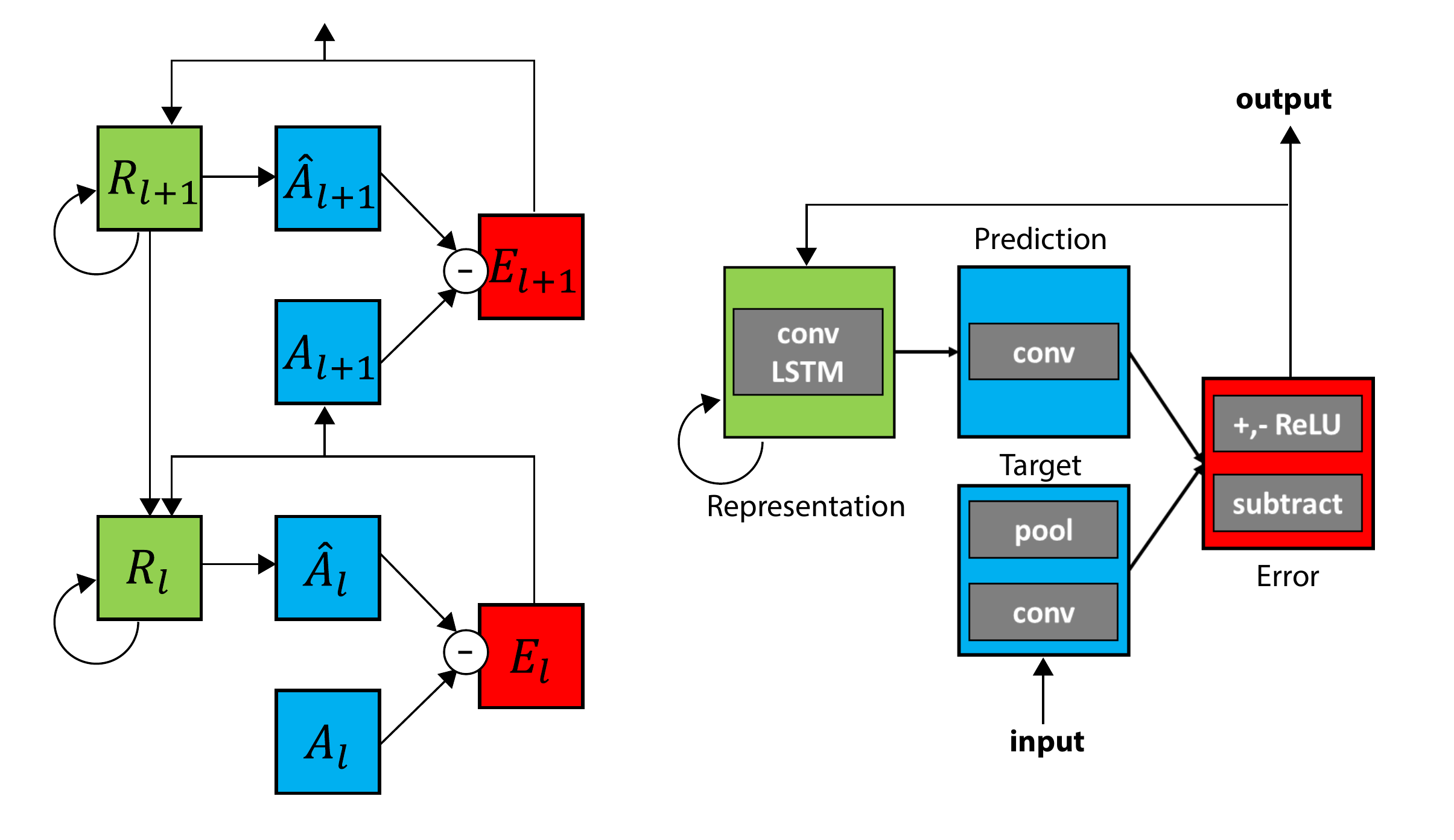}
  \end{center}
  \vspace{-9pt}
  \caption{Deep Predictive Coding Networks (PredNets) \cite{Lotter_2017}. Left: Each layer consists of representation neurons ($R_l$), which output a layer-specific prediction at each time step ($\hat{A}_l$), which is compared against a target ($A_l$) to produce an error term ($E_l$), which is then propagated laterally and vertically. Right: Module operations for case of video sequences. The target at the lowest layer of the network, $\hat{A}_0$, is set to the current input image.}
  \label{architecture}
\end{figure}

Here, inspired by past success in using ``out-of-the-box'' artificial deep neural networks as models of visual cortex, we explore whether modern predictive recurrent neural networks built for unsupervised learning can also explain dynamic phenomena in the brain.
In particular, we consider a deep predictive coding network (``PredNet''; \cite{Lotter_2017}), a network that learns to perform next-frame  prediction in video sequences \cite{Ranzato_2014, Brabandere_2016, Finn_2016, Mathieu_2015, SVVP, VPN, Vondrick_2017, Villegas_2017v2, Villegas_2017}.
The PredNet is motivated by the principle of ``predictive coding'' \cite{Rao_1999, Friston_2005, Spratling_2012, Chalasani_2013, Wen_2018}; the network continually generates predictions of future sensory data via a top-down path, and it sends prediction errors in its feedforward path (Fig.~\ref{architecture}).
At its lowest layer, the network predicts the input pixels at the next time-step, and it has been shown to make successful predictions in real-world settings (e.g. the KITTI car-mounted camera dataset \cite{Geiger2013IJRR}).
The internal representations learned from video prediction also proved to be useful for subsequent decoding of underlying latent parameters of the video sequence, consistent with the suggestion of prediction as a good loss function for unsupervised learning \cite{Softky_1996, Palm_2012, Lotter_2015, Wang_2015, Mathieu_2015, Srivastava_2015, OReilly_2014, Dosovitskiy_2017, Finn_2016}.

Predictive coding has a rich history in neuroscience literature \cite{Rao_2000, Summerfield_2006, Bastos_2012, Kanai_2015, Srinivasan_1982, Atick_1992}.
Rao and Ballard helped popularize the notion of predictive coding in neuroscience in 1999, proposing that spatial predictive coding could explain a key property of neurons in primary visual cortex (V1) known as end-stopping (\cite{Rao_1999}; see Section~\ref{end_stopping}).
Predictive coding has also been proposed as an explanatory framework for a variety of sensory systems in neuroscience \cite{Sukhbinder_2011, Zelano_2011, Mumford_1992}.
The PredNet formulates predictive coding principles in a deep learning framework to work on natural sequences, providing an opportunity to test a wide range of neuroscience phenomena using a single model.
Below, we show that despite being trained only to predict next frames in video sequences, the PredNet naturally captures a wide array of seemingly unrelated fundamental properties of neuronal responses and perception, including on/off dynamics, length suppression, sequence learning effects in visual cortex, norm-based coding of faces, illusory contours, and the flash-lag illusion.




\vspace{-2pt}
\section{Deep Predictive Coding Networks}

The deep predictive coding network proposed in \cite{Lotter_2017} (``PredNet'') consists of repeated, stacked modules where each module generates a prediction of its own feedforward inputs, computes errors between these predictions and the observed inputs, and then forwards these error signals to subsequent layers.
The model consists of four components: targets to be predicted ($A_l$), predictions ($\hat{A}_l$), errors between predictions and targets ($E_l$), and a recurrent representation from which predictions are made ($R_l$).
On an initial time step, the feedforward pass can be viewed as a standard CNN, consisting of alternating convolutional and pooling layers.
Predictions are made in a top-down pass via convolutions over the representational units, which are first updated using the representational units from the layer above and errors from the previous time step as inputs.
The $R_l$ units are implemented as convolutional LSTMs \cite{Hochreiter_1997, Shi_2015}.
Here, for the sake of biological interpretability, we replace the $tanh$ output activation function for the LSTMs with a $relu$ activation, enforcing positive ``firing rates''.
On the KITTI dataset this leads to a marginally ($8$\%) worse prediction mean-squared error (MSE) than the standard formulation, but it is still $2.6$ times better than the MSE that would be obtained by simply copying the last frame seen (compared to $2.8$ for $tanh$).

The error layers in the model, $E_l$, are calculated as a simple difference between the targets and predictions, followed by splitting into positive and negative error populations with $relu$ rectification.
The loss function for the network is set as the (weighted) sum of the error activations across each layer. 
We utilize the $L_{all}$ formulation presented in the original paper, which places a non-zero loss on the error unit activity in every level in the network.
Except where stated otherwise, results presented here use a model trained on the KITTI car-mounted camera dataset \cite{Geiger2013IJRR}.
The same model hyperparameters were used as presented in the paper (besides the $relu$ activation in the LSTM units).
Particularly, the model consists of four layers.
With $0$-indexing used here, Layer $1$ would be analogous to V1 in visual cortex.

\vspace{-3pt}

\section{Single Neuron Response Properties}

We begin by comparing the response properties of units in the PredNet to established single unit response properties of neurons in the primate visual system, which have been studied extensively using microelectrode recordings.
Here, we primarily compare response properties in the PredNet's error (``E'') units, the output units of each layer, to neuronal recordings in the superficial layers of cortex.
Response properties of other units in the PredNet (e.g. the ``R'' units) are included in the Supplemental Materials, and would likely map onto other parts of the cortical circuit. 


\paragraph{On/Off Temporal Dynamics}

\begin{figure}[h]
	\begin{center}
		\includegraphics[width=0.90\textwidth]{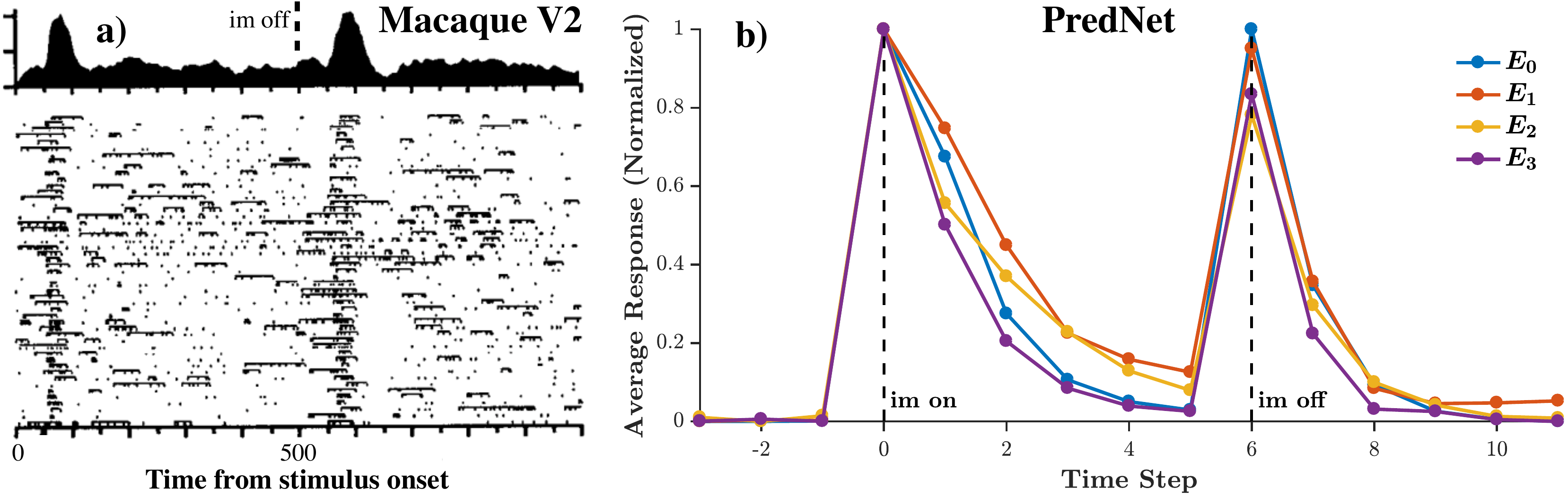}
	\end{center}
	\vspace{-9pt}
	\caption{On/off temporal dynamics. Left: Exemplar macaque V2 neuron responding to a static image. Reproduced with permission from Schmolesky et al. \cite{Schmolesky_1998}. Right: PredNet response to a set of naturalistic objects appearing on a gray background, after training on KITTI. Responses are grouped by layer for the $E$ units, and averaged across all units and all stimuli, per layer. }
	\vspace{-8pt}
	\label{on_off}
\end{figure}

As mentioned in the introduction, a conspicuous feature of visual cortical neuron responses is that they are highly dynamic, even when a static, unchanging image is presented to the subject. 
As an example of the commonly seen pattern of image on/off dynamics, Fig.~\ref{on_off}a shows a raster plot and peri-stimulus-time histogram of a recorded neuron in the secondary visual cortex (V2) of a macaque monkey \cite{Schmolesky_1998}. 
Peaks in firing rate shortly after image display and removal are prominent.
Fig.~\ref{on_off}b shows the average response of PredNet $E$ units in different layers over a set of $25$ naturalistic objects appearing on a gray background. 
The on/off dynamics are apparent on the population average level, for all four layers of the network.
These dynamics are also evident at the individual unit level, as illustrated in the Supplement, though there is variability.
While on/off dynamics have an an ``error''-like quality---an object unpredictably appears and disappears---these dynamics manifest in the $A$ and $R$ layers as well (see Supplement).

\paragraph{End-Stopping and Length Suppression}
\label{end_stopping}

\begin{figure}[h]
	\begin{center}
		\includegraphics[width=1.0\textwidth]{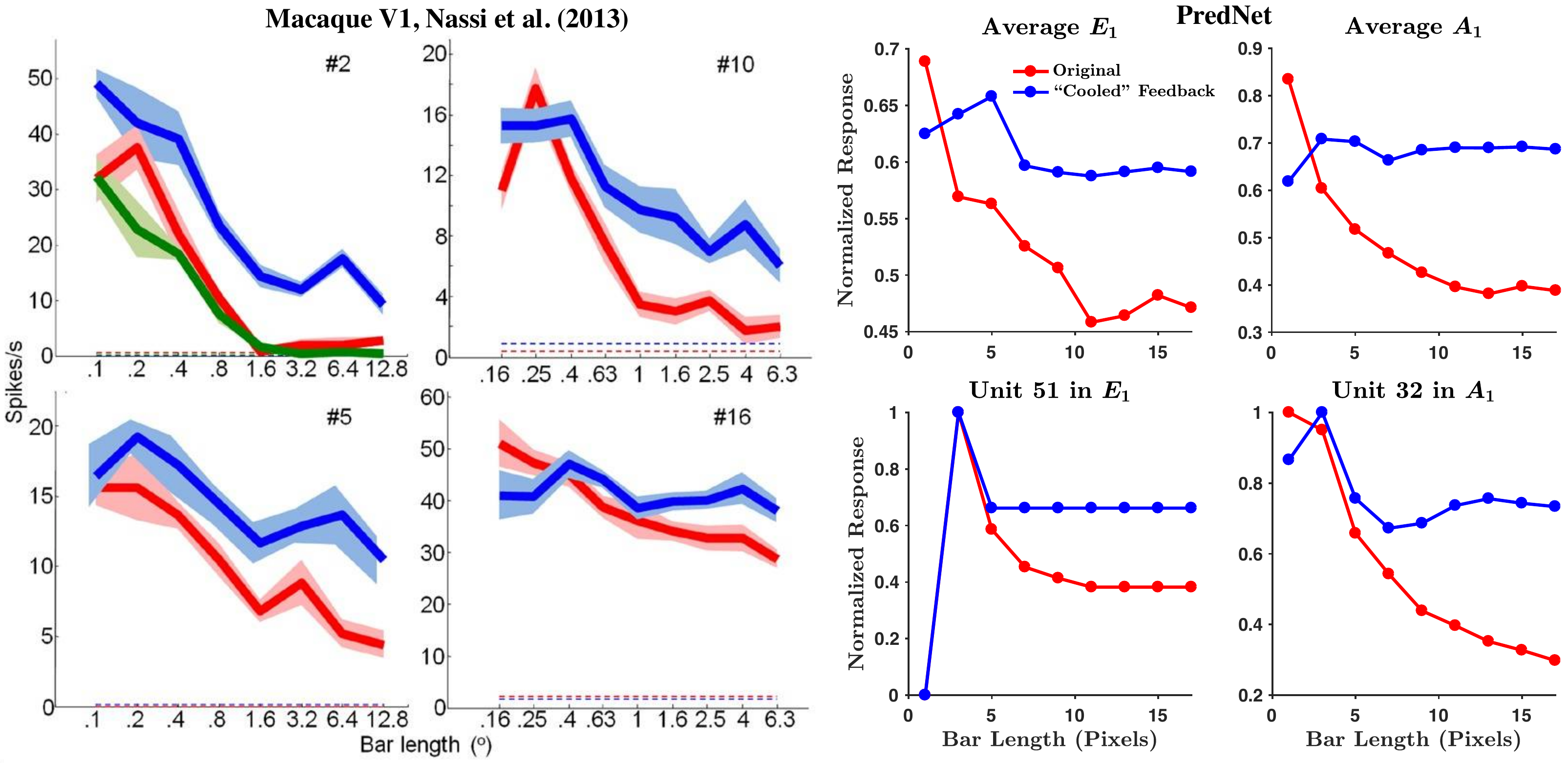}
	\end{center}
	\vspace{-9pt}
	\caption{Length suppression. Left: Responses of example macaque V1 units to bars of different lengths before (red), during (blue), and after (green) inactivation of V2 via cryoloop cooling. Reproduced with permission from \cite{Nassi_2013}. Right: PredNet after training on the KITTI dataset -- average $E_1$ and $A_1$, and examples. Red: Original network. Blue: Feedback weights from $R_2$ to $R_1$ set to zero.}
	\label{length_suppression}
\end{figure}

Prediction in time and prediction in space are inextricably intertwined.
As Rao and Ballard \cite{Rao_1999} illustrated, end-stopping in V1 can be explained by spatial predictive coding.
End-stopping, or length suppression, is the phenomenon where a neuron tuned for a particular orientation becomes less responsive to a bar at this orientation, when the bar extends beyond its classical receptive field \cite{Hubel_1968}.
The predictive coding explanation is that lines/edges tend to be continuous in nature, and thus the center of a long bar can be predicted from its flanks. 
A short, discontinuous bar, however, deviates from natural statistics, and responding neurons signal the deviation.
One potential source for conveying the long range predictions in the case of an extended bar could be feedback from higher visual areas with larger receptive fields.
This hypothesis was elegantly tested in Nassi et al. \cite{Nassi_2013} using reversible inactivation of V2 paired with V1 recordings in the macaque.
As illustrated in the left side of Fig.~\ref{length_suppression}, cryoloop cooling of V2 led to a significant reduction in length suppression, indicating that feedback from V2 to V1 is essential for the effect.

The right side of Fig.~\ref{length_suppression} demonstrates that length suppression, and its mediation through top-down feedback, are also present in the PredNet.
The upper left and right panels contain the mean normalized response for units in the $E_1$ and $A_1$ layers, respectively, to bars of different lengths.
The red curves correspond to the original network (trained on KITTI) and the blue curves correspond to zero-ing the feedback from $R_2$ to $R_1$. 
For each filter channel, a set of 2D Gabor wavelet stimuli was first used to determine the optimal orientation.
Responses to bars of different length at this orientation were then measured, as a sum of the activations over stimulus duration ($10$ time steps).
The bottom row contains exemplar $E_1$ and $A_1$ units.
Quantifying percent length suppression (\%LS) as $100*\frac{R_{max} - R_{longest \: bar}}{R_{max}}$, the median decrease in \%LS upon removing top-down signaling was $16$\% for $E_1$ units ($p < 0.05$, Wilcoxon signed rank test) and $33$\% for $A_1$ units ($p < 0.0005$).
For $R_1$ units, the median \%LS decrease was $2$\% ($p=0.18$).
Indeed, the average $R_1$ response did not exhibit much length suppression (see Supplement), though, there were particular examples with a strong effect. 
\vspace{-3pt}
\paragraph{Sequence Learning Effects in Visual Cortex}

\begin{figure}[h]
	\begin{center}
		\includegraphics[width=0.9\textwidth]{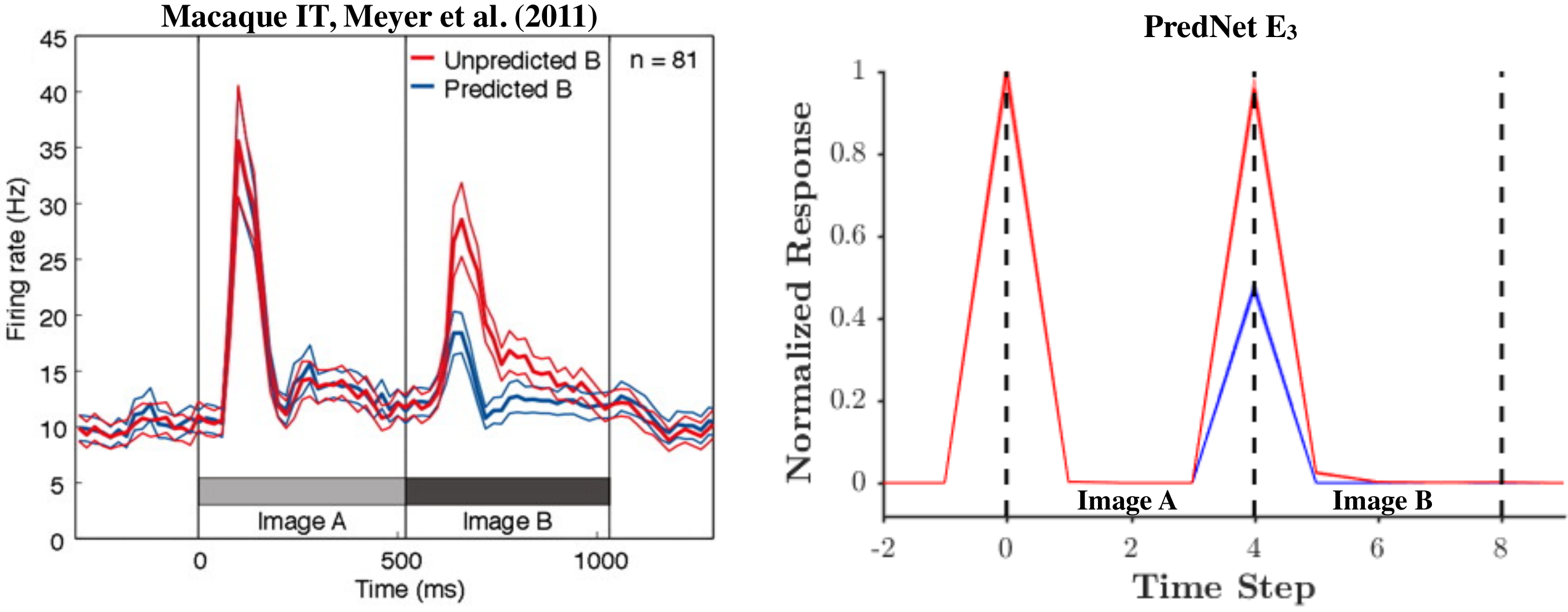}
	\end{center}
	\vspace{-12pt}
	\caption{Predicted vs. unpredicted image transitions. Left: Mean of $81$ neurons recorded in macaque (IT). Reproduced with permission from \cite{Meyer_2011}. Right: Mean ($\pm$ SE) across PredNet $E_3$ units.}
	\label{image_pairing}
\end{figure}

Predictions are often informed by recent experience, and violations of these predictions can be highly salient. 
Meyer and Olson \cite{Meyer_2011} provided a striking example of this in visual cortex via image sequence learning.
The authors exposed monkeys to image pairs in a fixed order for over $800$ trials for each pair.
The left panel of Fig.~\ref{image_pairing} shows the mean response of $81$ IT neurons in a subsequent testing period, for predicted and unpredicted pairs.
When the second image differs from expectations, the response is much stronger than when the expected image is presented.

\vspace{-5pt}
\begin{figure}[h]
\vspace{-3pt}
	\begin{center}
		\includegraphics[width=0.9\textwidth]{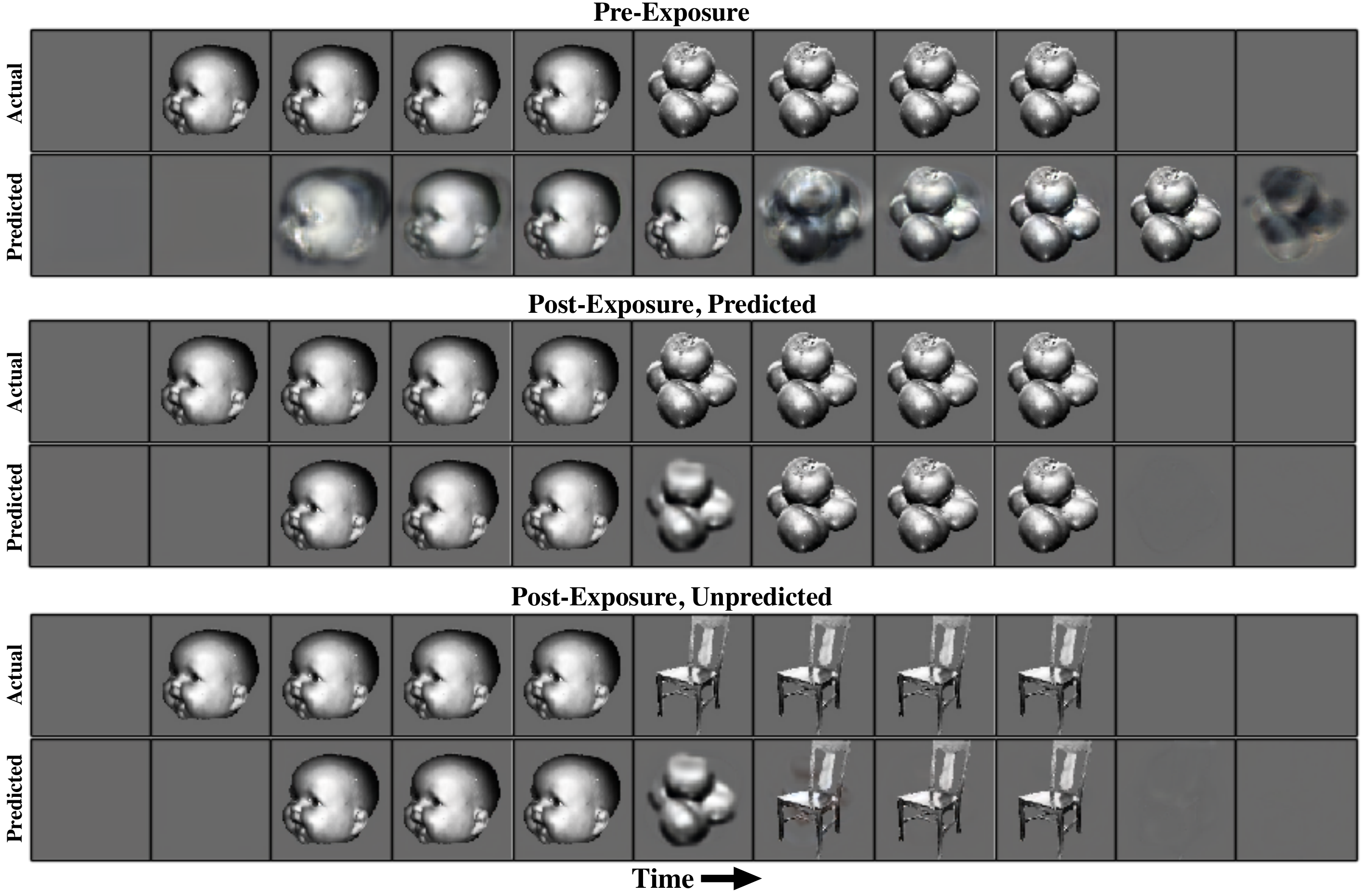}
	\end{center}
	\vspace{-11pt}
	\caption{Learned image transitions. Top: Predictions of a KITTI-trained PredNet model on an example sequence. Middle: PredNet predictions after repeated ``training'' on the sequence. Bottom: PredNet predictions for an unpredicted image transition.}
	\label{image_pairing_examples}
\end{figure}

The right panel of Fig.~\ref{image_pairing} demonstrates a similar effect in the PredNet after an analogous experiment.
The model was trained on five image pairs for $800$ epochs. 
Fig.~\ref{image_pairing_examples} contains an example sequence and the corresponding next-frame predictions before and after the training.
The model, prior to exposure to the images in this experiment (trained only on KITTI), settles into a noisy, copy-last-frame prediction mode.
After exposure, the model is able to successfully make predictions for the expected image pair (row $2$).
Since the chosen image pair is unknown \emph{a priori} and the model is fully convolutional, the initial prediction is the constant gray background when the first image appears.
The model then rapidly copies this image for the ensuing three frames.
Next, the model successfully predicts the transition to the second object (a stack of tomatoes in this case).
In row $3$, a sequence that differs from the training pair is presented.
The model still makes the prediction of a transition to tomatoes, even though a chair is presented, but then copies the chair into subsequent predictions.
Fig.~\ref{image_pairing} shows that the unexpected transitions result in a larger average response in the final $E$ layer of the network.
In fact, in all levels and all unit types ($E$, $A$, $R$), there is a larger response to the unpredicted images (Supp. Table 1).
The overall magnitude of the difference is similar for $A$ and $E$, and is lower for $R$.

\paragraph{Norm-Based Coding of Faces}

\begin{figure}[h]
	\begin{center}
		\includegraphics[width=0.9\textwidth]{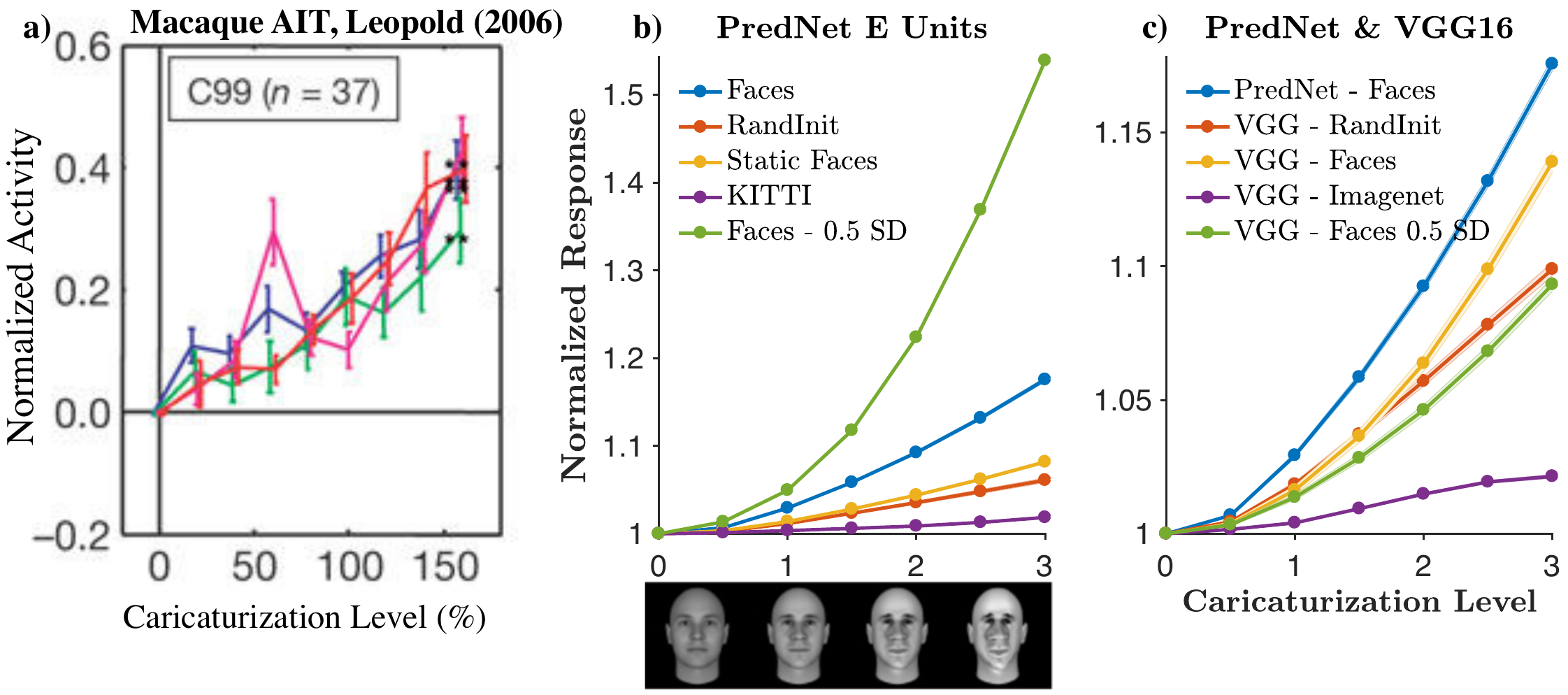}
	\end{center}
	\vspace{-11pt}
	\caption{Norm-based coding of faces. a): Population response of $37$ neurons recorded in macaque IT for four different faces, as a function of caricaturization. Reproduced with permission from \cite{Leopold_2006}. b): Mean PredNet $E$ unit responses for models trained on different stimuli. Faces - Rotating synthetic faces \cite{Lotter_2017}. RandInit - Random initial weights. Static Faces - Same collection of images used in Rotating Faces except presented statically. Faces 0.5 SD - Rotating faces except each face generated from a distribution with half of the original standard deviation. c): Comparing PredNet with VGG16. VGG Faces - VGG trained in a siamese manner on a same/different task using the static face images.}
	\vspace{-3pt}
	\label{norm_faces}
\end{figure}

The representation of deviations from expectations can also explain observed neural embeddings of familiar stimuli such as faces.
The norm-based coding theory suggests that faces are encoded with respect to a mean face \cite{Rhodes_2006}.
Leopold et al. \cite{Leopold_2006} demonstrated that face-responsive neurons in macaque anterior IT are frequently tuned monotonically (often positively) to directions away from an average face (Fig.~\ref{norm_faces}).
This was tested by using synthetic faces with continuously varying levels of caricature. 
Training for next-frame prediction of rotating, computer-generated faces \cite{Lotter_2017}, we see similar effects in the PredNet.
The faces were created using software implementing a principal component analysis of a corpus of human faces (FaceGen \cite{facegen}).
For training, $16K$ sequences were generated with a random face, initial orientation, and rotation velocity.
The blue curve in Fig.~\ref{norm_faces}b illustrates the post-training response of $E$ units in the network as a function of caricaturization level for $200$ previously unseen faces.
The response is calculated by first averaging the response of all units in a given layer, then averaging the layer responses.
The PredNet $E$ units become significantly more responsive to increasing levels of caricature after predictive training compared to the random initialization (red curve).
This effect is diminished when the same network is trained on the same set of images, but in a static, autoencoder fashion (yellow curve).
Note that the randomly initialized network already responds more highly to caricature, likely because the caricature faces tend to have higher contrast, sharper edges, etc., which even random CNNs can be tuned for \cite{Saxe_2010}.
When training on an unrelated dataset (e.g. the KITTI dataset), this effect is reduced (purple curve).
Training the network on rotating faces that had been generated using half the standard deviation for each principal component results in an even larger caricature response (green curve).
All of these effects are consistent in the $A$ units as well, though the results are mixed for $R$ (Supplement).

Fig.~\ref{norm_faces}c compares the PredNet responses to a popular CNN, VGG16 \cite{Simonyan_2015}.
VGG responses were quantified by averaging over the outputs of its five convolutional blocks (after the max-pooling).
Similar to the PredNet, the randomly initialized VGG16 displayed higher activity for more caricatured faces.
ImageNet training decreased this effect, akin to the PredNet KITTI training.
To train VGG on the synthetic faces, a same/different task was performed using the network in a siamese fashion.
Using the same images as the PredNet, a training example consisted of a pair of images at different orientations with a binary cross-entropy objective for same/different identity classification.
This training procedure resulted in an increased response to higher caricature levels (at least on the original dataset -- yellow curve in Fig.~\ref{norm_faces}c), although somewhat less than the PredNet $E$ units.
While analogous norm-based face encoding effects can be seen with discriminatively-trained (VGG) models, the PredNet architecture is able to capture these same effects (even more strongly) in a fully unsupervised way.

\section{Visual Illusions}

Visual illusions can provide powerful insight into the underpinnings of perception.
Here we demonstrate that the PredNet exhibits correlates of two illusions: illusory contours and the flash-lag effect, both of which have aspects of spatial and temporal prediction.
The PredNet has also recently been shown to predict the illusory motion perceived in the rotating snakes illusion \cite{Watanabe_2018}.

\vspace{-3pt}
\paragraph{Illusory Contours}

\begin{figure}[h]
	\begin{center}
		\includegraphics[width=1.0\textwidth]{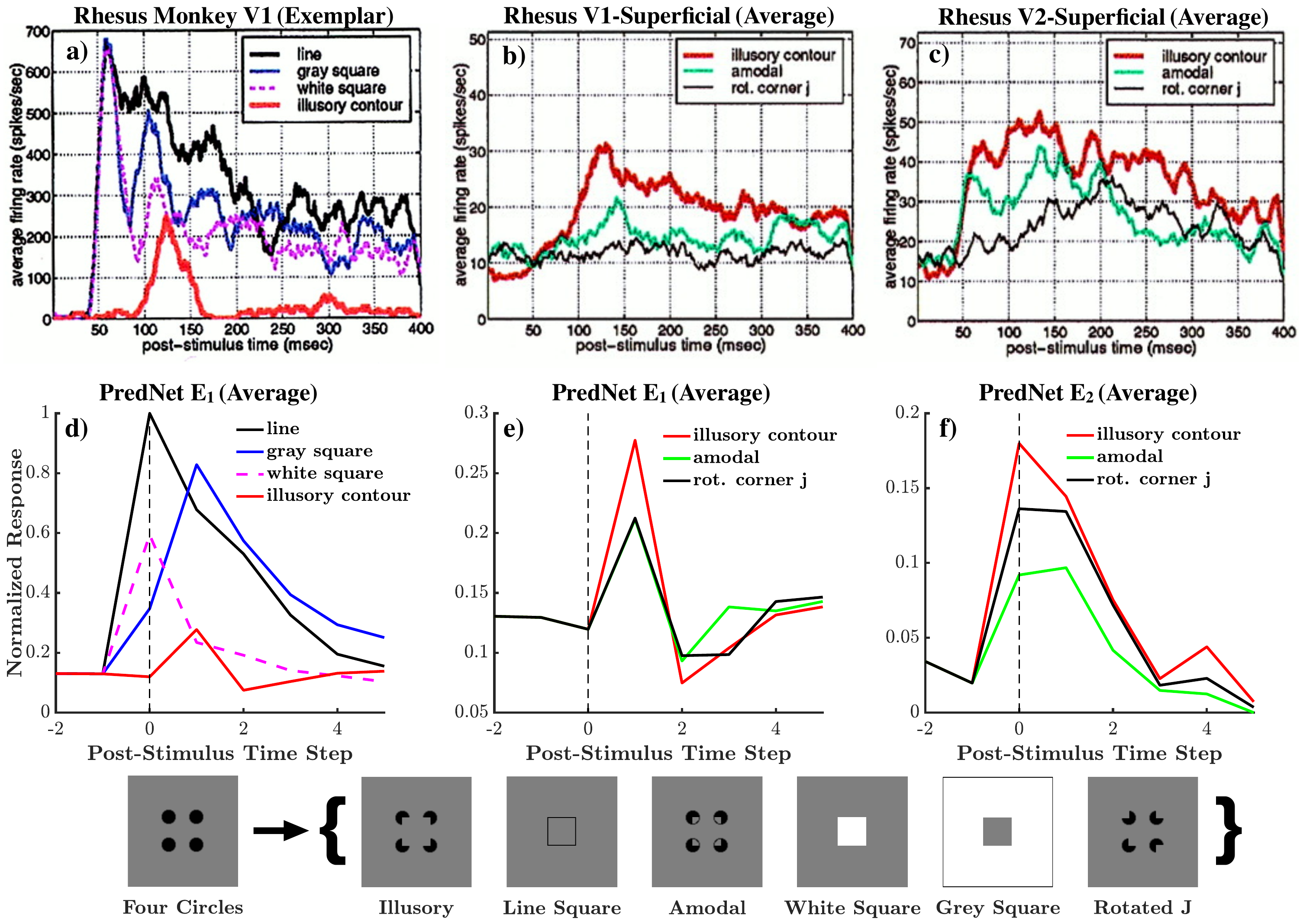}
	\end{center}
	\vspace{-9pt}
	\caption{Illusory contours. Top: Reproduced with permission from Lee et al. \cite{Lee_2001}. A given trial consisted of the four circles image abruptly transitioning to one of the displayed test images.}
	\label{illusory_contours}
\end{figure}

Illusory contours, as in the Kanizsa figures \cite{Kaniza}, elicit perceptions of edges and shapes, despite the lack of enclosing lines.
Lee et al. \cite{Lee_2001} found that neurons in monkey V1 can be responsive to illusory contours, albeit at a reduced response and increased latency to physical contours.
Fig.~\ref{illusory_contours}a contains an example of such a neuron. 
The stimuli in the experiment consisted of sequences starting with an image of four circles, which then abruptly transitioned to one of numerous test images, including the illusion.
Illustrated in Fig.~\ref{illusory_contours}b, the population average of $49$ superficial V1 neurons responded more strongly to the illusion than similar, but non-illusory stimuli.
This preference was also apparent in V2, with a response that was, interestingly, of a shorter latency compared to V1.

Fig.~\ref{illusory_contours}c-e demonstrate that the effects discovered by \cite{Lee_2001} are also present in the PredNet.
In the population average of $E_1$ units, there is indeed a response to the illusory contour, and at an increased latency compared to a physical line square (Fig.~\ref{illusory_contours}c).
The response was calculated separately for each filter channel, by first finding the optimal orientation using short bar segments.
The responses were then normalized (division by max response over all stimuli) and averaged.
Fig.~\ref{illusory_contours}d illustrates that the average $E_1$ response was higher for the illusory contour than the similar control images.
This was also the case for $E_2$ units, with a peak response one time step before $E_1$.
Indeed, the size of the stimuli was such that it was larger than the feedforward receptive field of the layer $1$ neurons, but smaller than that of the layer $2$ neurons (matching the protocol of \cite{Lee_2001}).
Quantifying the preference of the illusion to the amodal and rotated ``J'' images for each individual unit \cite{Lee_2001}), we find that the average is positive (more preference to the illusion) for all tested layers ($E$, $A$, $R$, layers 1,2).


\vspace{-3pt}
\paragraph{The Flash-Lag Effect}

\begin{figure}[h]
\vspace{-3pt}
	\begin{center}
		\includegraphics[width=0.8\textwidth]{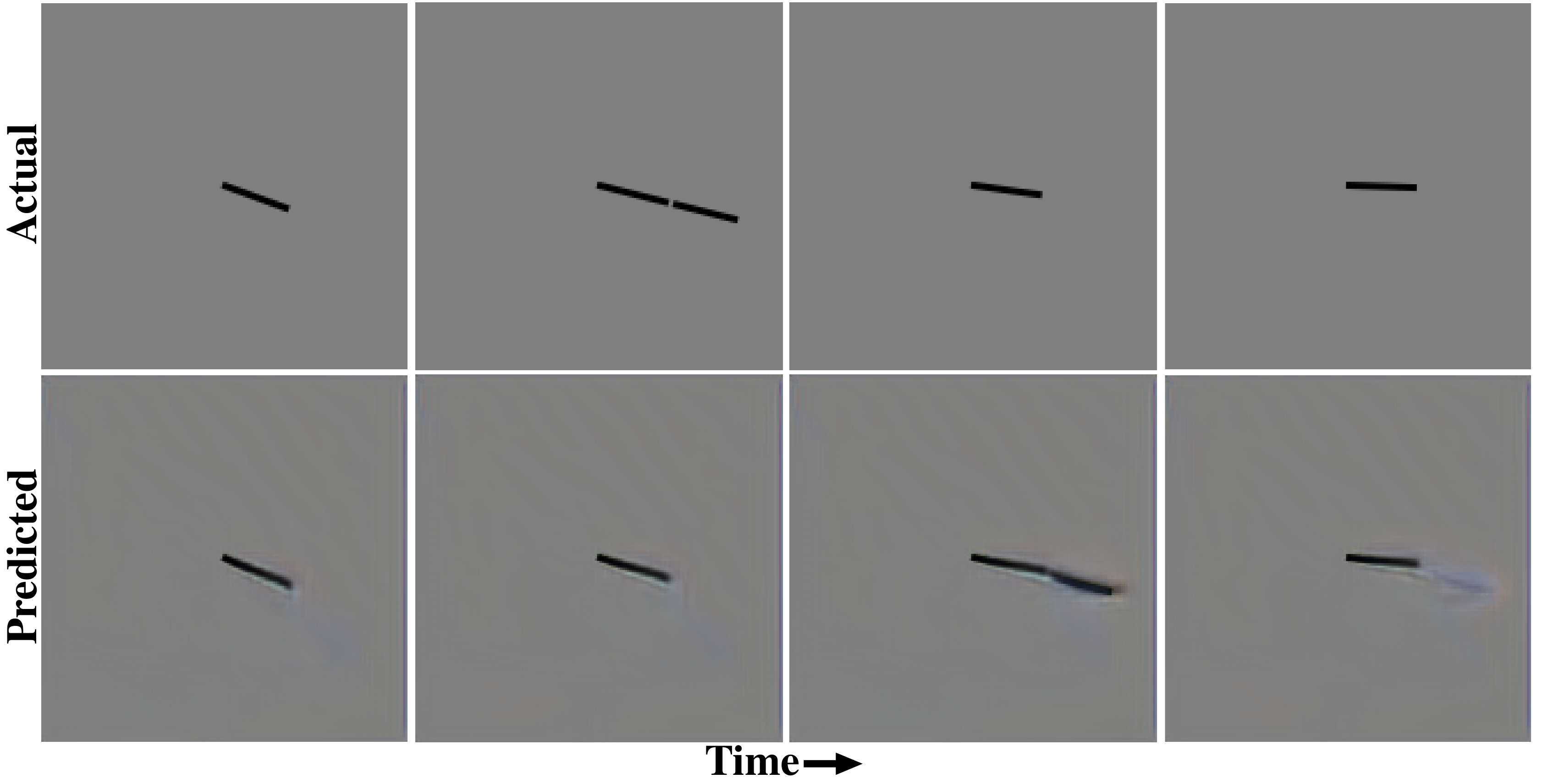}
	\end{center}
	\vspace{-13pt}
	\caption{Flash-lag effect. Top: Stimulus clip. Bottom: PredNet predictions after KITTI training.}
	\label{flash_lag_example}
\end{figure}

Another illusion for which prediction has been proposed as having a role is the flash-lag effect.
Fundamentally, the flash-lag effect describes illusions where an unpredictably appearing stimulus (e.g. a line or dot) is perceived as ``lagging'' a predictably moving stimulus nearby, even when the stimuli are, in fact, precisely aligned in space.
These illusions are sometimes interpreted as evidence that the brain is performing inference to predict the likely true current position of a stimulus, even in spite of substantial latency (up to hundreds of milliseconds) in the visual system \cite{Khoei_2017}.
The version of the illusion tested here consists of an inner, continuously rotating bar and an outer bar that periodically flashes on.
Fig.~\ref{flash_lag_example} contains an example prediction by the PredNet on a sample sequence within a flash-lag stimulus.
The rotation speed of the inner bar in the clip was set to $6$ degrees per time step.
The first feature of note is that the PredNet is indeed able to make reasonable predictions for the inner rotating bar.
Quantifying this, the average angle of the bar in the outputted predictions is 1.4\rpm1.2\degree (s.d.) behind the actual bar (see Supp. Methods), which is significantly less than a $6$\degree difference, which would result from simply copying the last seen frame.
Again, the model was trained on real-world videos, so the generalization to this impoverished stimulus is non-trivial.
Secondly, the post-flash predictions made by the model tend to resemble the perceived illusion.
The average angular difference between the predicted outer bar and inner bar is 6.8\rpm2.0\degree.
Considering that the model was trained for next frame prediction on a corpus of natural videos, this suggests that our percept matches the statistically predicted next frame (as estimated by the PredNet) more than the actual next frame.
This natural statistics interpretation of the flash-lag illusion has, in fact, been similarly suggested by Wojtach et al. \cite{Wojtach_2008}.

\vspace{-5pt}
\section{Discussion}

We have shown that an off-the-shelf recurrent neural network trained to predict future video frames can explain a wide variety of seemingly unrelated phenomena observed in visual cortex and visual perception.
These phenomena range from the details of responses of individual neurons, to complex visual illusions. 
Importantly, throughout, we used a base model trained on natural videos. 
Our work adds to a growing body of literature showing that deep neural networks trained to perform relevant tasks can serve as surprisingly good models of biological neural networks, often even outperforming models designed to explain neuroscience phenomena.

While we have shown that the PredNet architecture demonstrates a wide range of phenomena reminiscent of biology, we do not claim that the PredNet architecture \emph{per se} is required to explain these phenomena.
Rather, we argue that the network is \emph{sufficient} to produce these phenomena, and we note that explicit representation of prediction errors in units within the feedforward path of the PredNet provides a straightforward explanation for the transient nature of responses in visual cortex in response to static images.
That a single, simple objective---prediction---can produce such a wide variety of observed neural phenomena underscores the idea that prediction may be a central organizing principle in the brain \cite{Rao_1999}, and points toward fruitful directions for future study in both neuroscience and machine learning.

\subsubsection*{Acknowledgments}
This work was supported by IARPA (contract D16PC00002), the National Science Foundation (NSF IIS 1409097), and the Center for Brains, Minds and Machines (CBMM, NSF STC award CCF-1231216).

\bibliography{main.bib}

\bibliographystyle{ieee}

\clearpage
\section{Supplementary Material}
\subsection{On/Off Temporal Dynamics}

Temporal dynamics were tested with a set of $25$ objects.
Examples of the objects can be seen in Fig.~\ref{image_pairing_examples}.
Testing sequences consisted of a gray background for $7$ time steps, followed by an object on the background for $6$ time steps.
As a general theme, we see some diversity in the response profiles of all units, but especially those in the ``R'' layers.
We have focused in the main text on the ``E'' units, which most naturally map onto Layer 2/3 cortical pyramidal neurons (which are the output units in a putative cortical microcircuit).
Diversity of responses is also observed throughout the neuroscience literature, and we hypothesize that to the extent that units in the PredNet map in a direct way onto cortical circuits \cite{Bastos_2012}, the less-often experimentally sampled deep neurons might be reasonable analogs to the ``R'' units.
We present representative units from all parts of the PredNet here for completeness.

\begin{figure}[h]
	\begin{center}
		\includegraphics[width=1.0\textwidth]{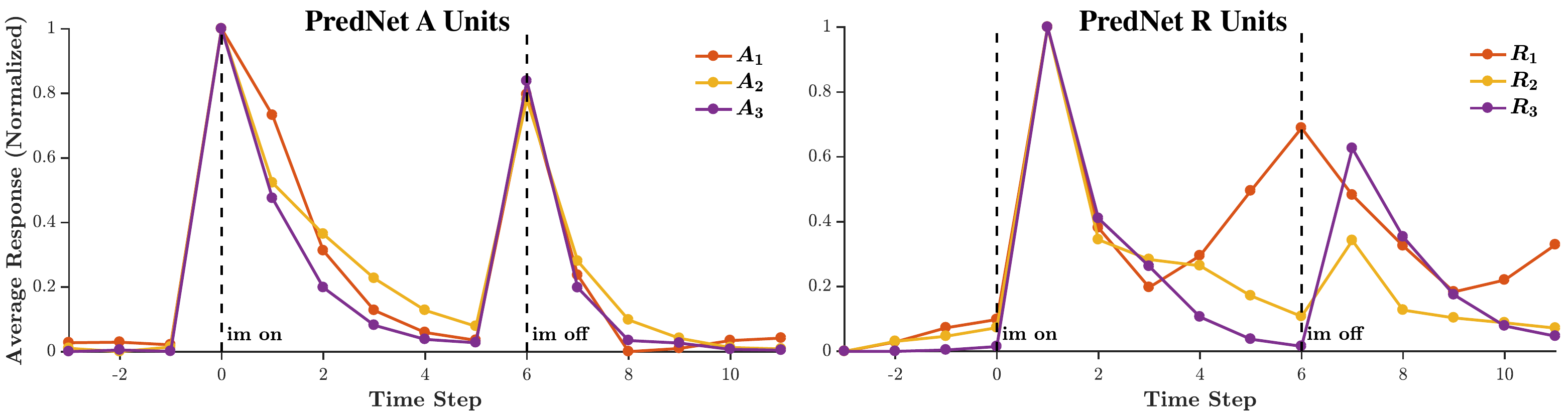}
	\end{center}
	\caption{Average temporal dynamics in $A$ and $R$ units in response to set of naturalistic objects on a gray background, after training on the KITTI dataset.}
	\label{on_off_AR}
\end{figure}

Summary responses for the $A$ and $R$ units are contained in Fig. \ref{on_off_AR}.
The responses are grouped per layer and consist of an average across all the units (all filters and spatial locations) in a layer.
The mean responses were then normalized between $0$ and $1$.
Responses for layer $0$, the pixel layer, are omitted in Fig. \ref{on_off_AR} because of their heavy dependence on the input pixels for the $A$ and $R$ layers.
Note that, by notation in the network's update rules, the input image reaches the $R$ layers at a time step after the $E$ and $A$ layers.

As illustrated in Fig. \ref{on_off_AR}, the $A$ and $R$ layers seem to generally exhibit on/off dynamics, similar to the $E$ layers.
$R_1$ also seems to have another mode in its response, specifically a ramp up between time steps $3$ and $5$ post image onset.
As will be illustrated below, this results from a few strongly firing neurons that exhibit this pattern.

\begin{figure}[h]
	\begin{center}
		\includegraphics[width=1.0\textwidth]{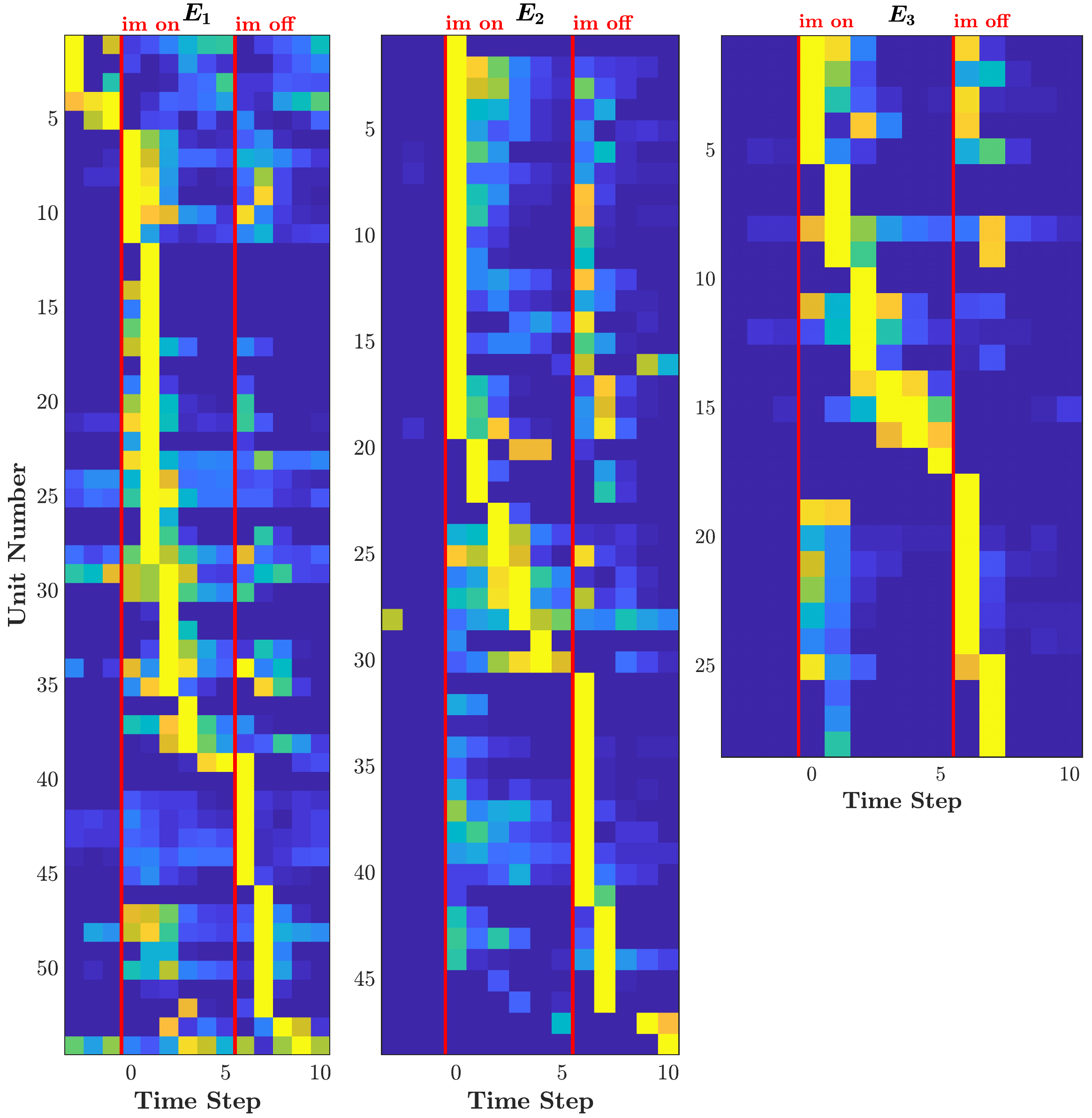}
	\end{center}
	\caption{Mean response of each active unit in the $E$ layer at the central receptive field. Each row (unit) is normalized by its max response.}
	\label{E_raster}
\end{figure}

\begin{figure}[h]
	\begin{center}
		\includegraphics[width=1.0\textwidth]{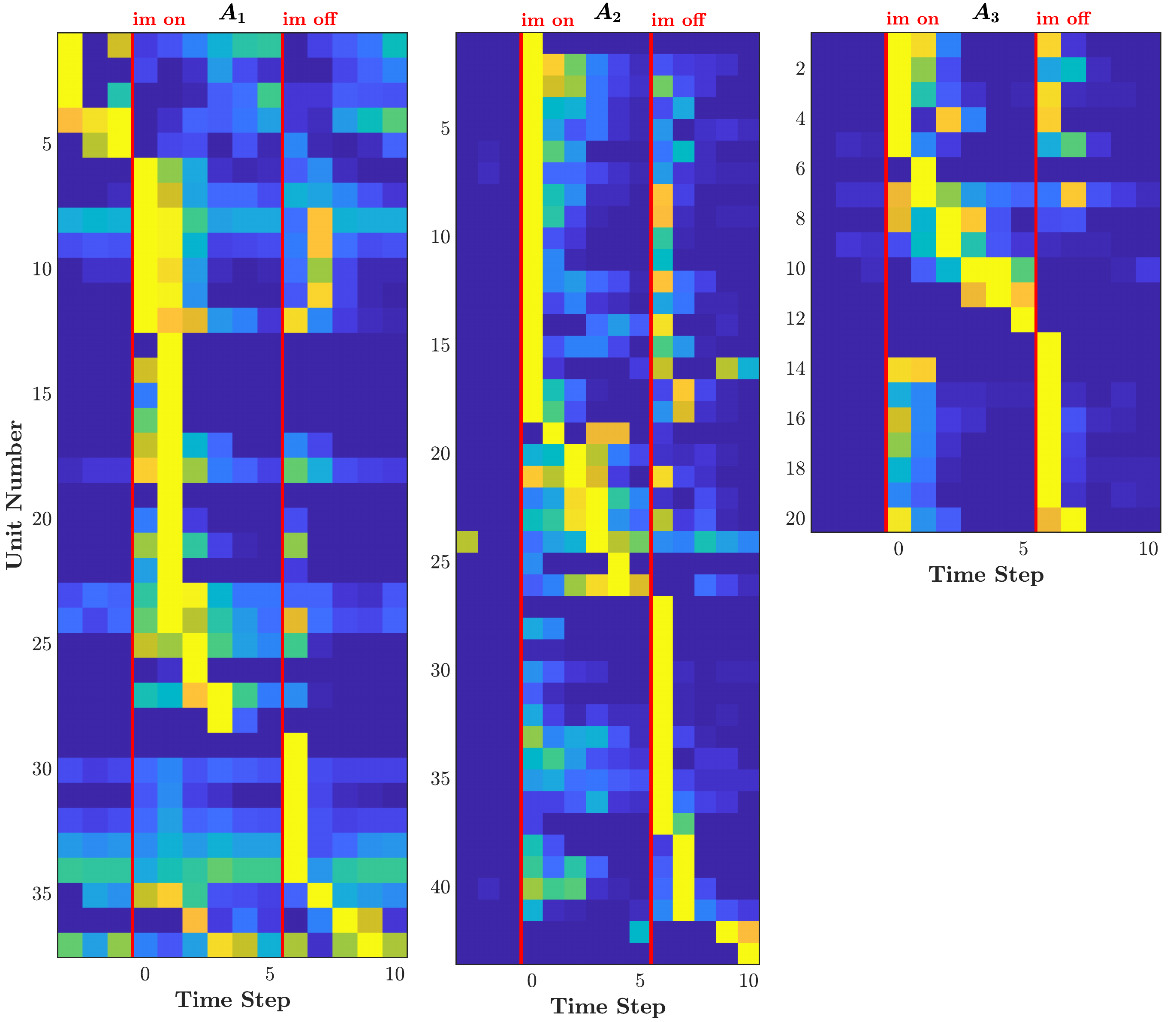}
	\end{center}
	\caption{Mean response of each active unit in the $A$ layer at the central receptive field. Each row (unit) is normalized by its max response.}
	\label{A_raster}
\end{figure}

\begin{figure}[h]
	\begin{center}
		\includegraphics[width=0.9\textwidth]{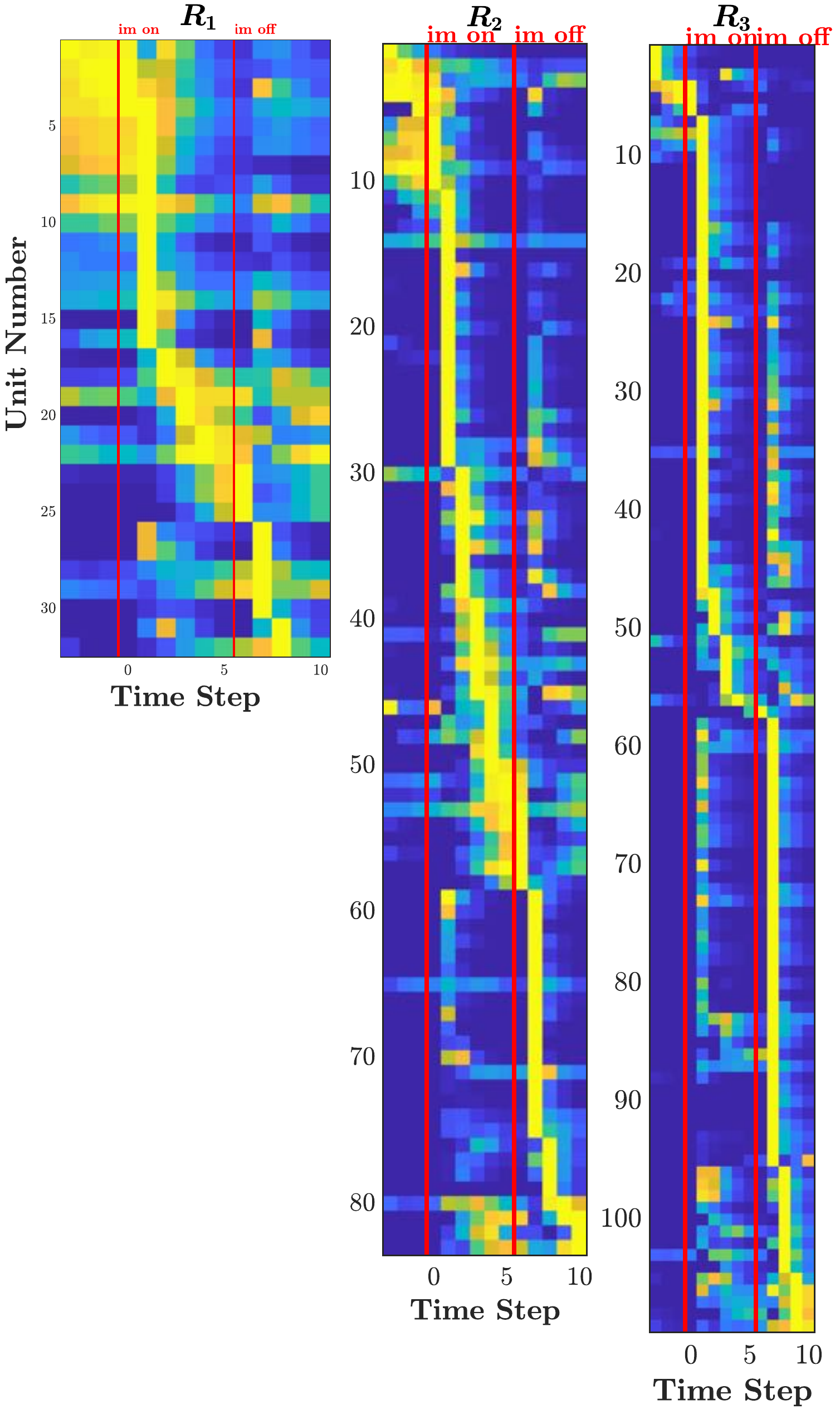}
	\end{center}
	\vspace{-9pt}
	\caption{Mean response of each active unit in the $R$ layer at the central receptive field. Each row (unit) is normalized by its max response.}
	\label{R_raster}
\end{figure}

Fig. \ref{E_raster}-\ref{R_raster} illustrate the variety of responses present among the units in the model.
Each plot for a given layer shows all the active units in the layer at the central receptive field.
The average response for each unit over the $25$ images is shown, and each row is normalized to have a max of one.
In each of the plots, it is apparent that a large proportion of the neurons have a peak response closely following image onset and/or offset.
However, there are a good number of neurons that have peaks at different times.
Overall, the on and off responses for individual neurons tend to be asymmetric -- some neurons have stronger ``on'' responses, some have stronger ``off'' responses.

\clearpage

\clearpage

\clearpage

\begin{figure}[h]
	\begin{center}
		\includegraphics[width=1.0\textwidth]{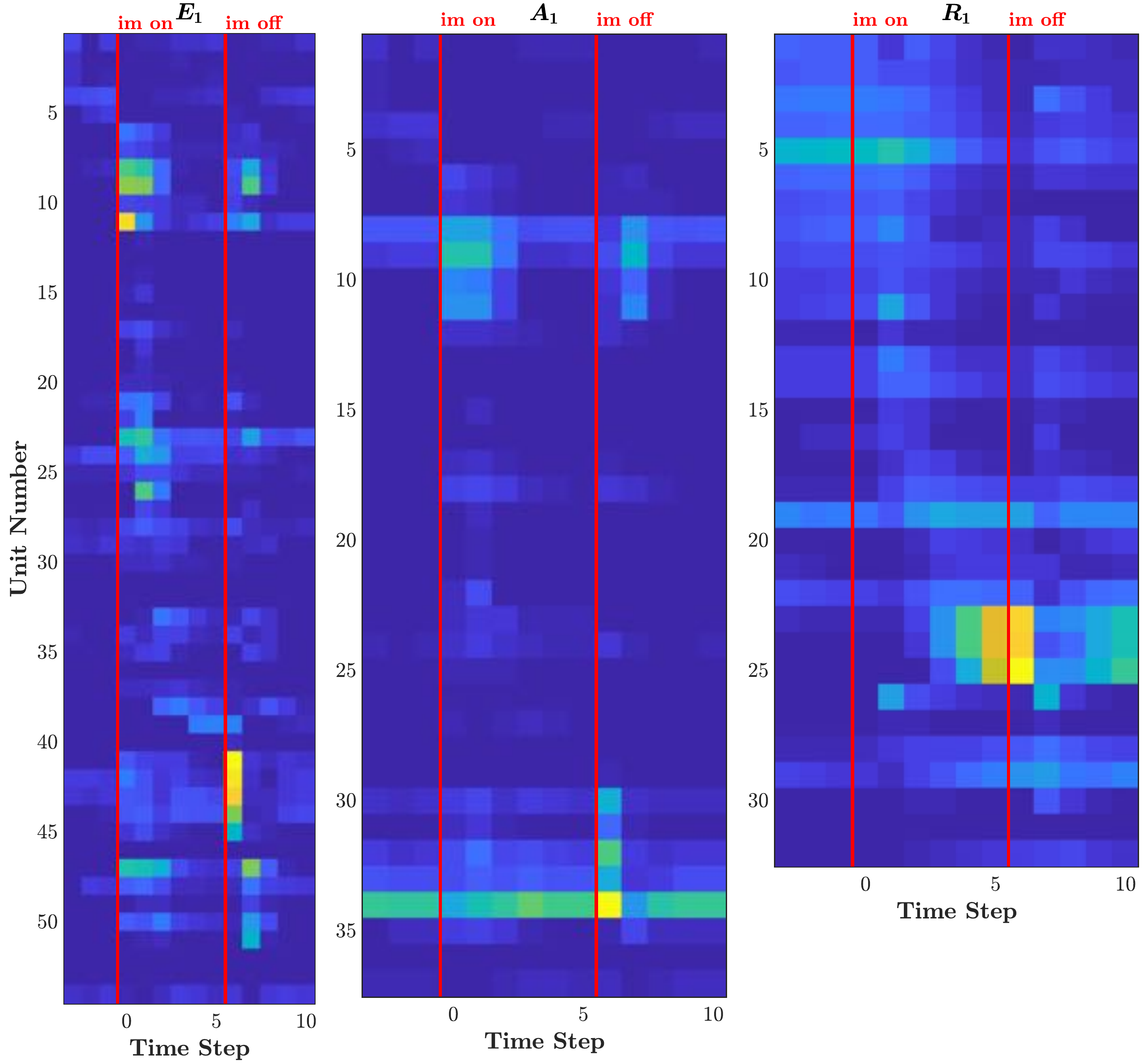}
	\end{center}
	\caption{Mean response of each active unit with a central receptive field for all three unit types at Layer $1$. Responses are normalized globally per each unit type.}
	\label{Layer1_raster}
\end{figure}

Figure \ref{Layer1_raster} is similar to the previous plots, except a global normalization is used instead of row normalization.
There are subsets of neurons that are particularly more active than others.
For $R_1$, rows $23$-$25$ contain examples of units that contribute to the ramping behavior from time-steps $3$ to $5$ in Fig. \ref{on_off_AR}.

\clearpage

\subsection{End-Stopping and Length Suppression}

\begin{figure}[h]
	\begin{center}
		\includegraphics[width=1.0\textwidth]{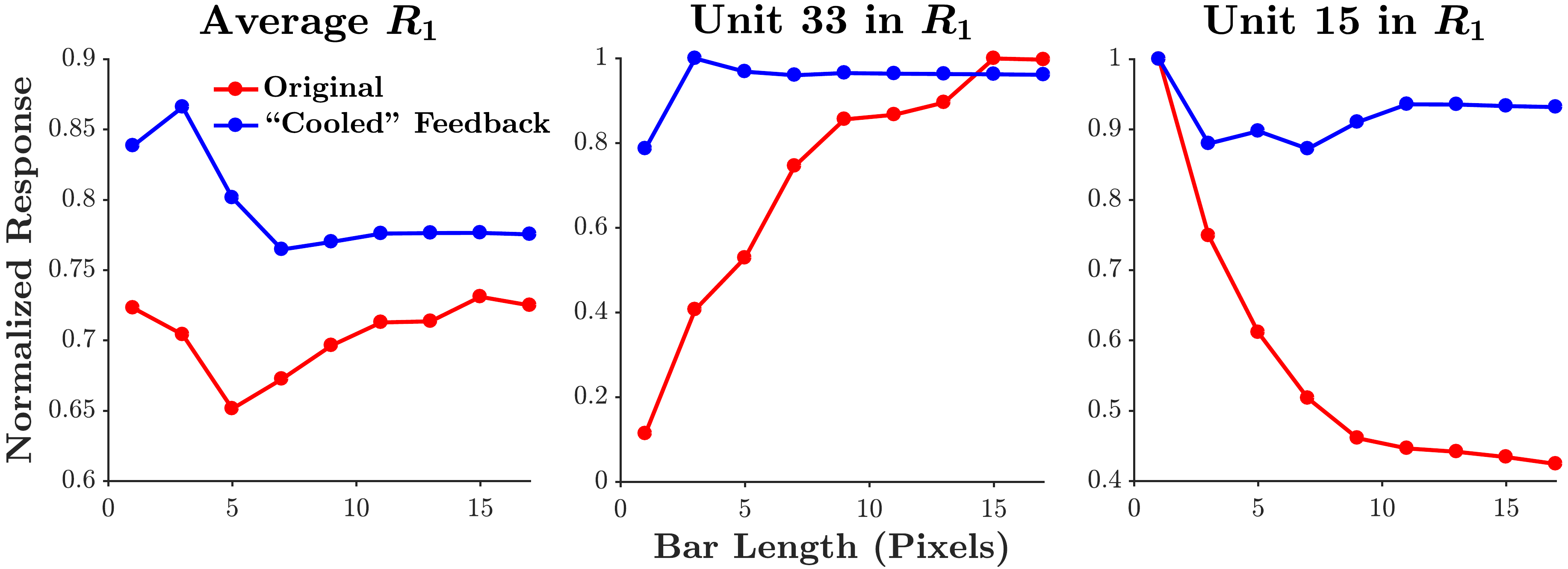}
	\end{center}
	\caption{Length suppression analysis for $R_1$ units.}
	\label{end_stopping_R}
\end{figure}

Responses to bars of different length were quantified for the length suppression experiment as a sum over the stimulus duration of $10$ time steps. 
The bars appeared on a gray background, which was first presented to the network for $5$ time steps, to allow the network to settle to a steady state before stimulus presentation.
For each filter channel, the response at the central receptive field was quantified and normalized to unit maximum before averaging.
The average $R_1$ response and two exemplar units are displayed in Fig. \ref{end_stopping_R}.
As mentioned in the main text, $R_1$ did not have a significant length suppression effect, with some neurons showing length suppression (right panel) and others showing an opposite effect (middle panel).

\subsection{Sequence Learning Effects in Visual Cortex}

For the exposure training phase in the learned sequence experiment, the Adam \cite{Kingma_2014} optimizer was used with default parameters.
Table~\ref{sequence_table} contains the percent increase in response between predicted and unpredicted sequences for each layer.

\begin{table}[h]
  \caption{Percent increase of response between predicted and unpredicted sequences}
  \label{sequence_table}
  \centering
  \begin{tabular}{ccccc}
    \toprule
    Unit Type     & Layer $0$     & Layer $1$   & Layer $2$ & Layer $3$ \\
    \midrule
    E   & $308$   & $90$    & $109$   & $108$     \\
    A   & N/A   & $78$    & $109$ & $108$       \\
    R   & N/A   & $18$  & $19$  & $30$  \\
    \bottomrule
  \end{tabular}
\end{table}

\subsection{Norm-Based Coding of Faces}

\begin{figure}[h]
	\begin{center}
		\includegraphics[width=0.8\textwidth]{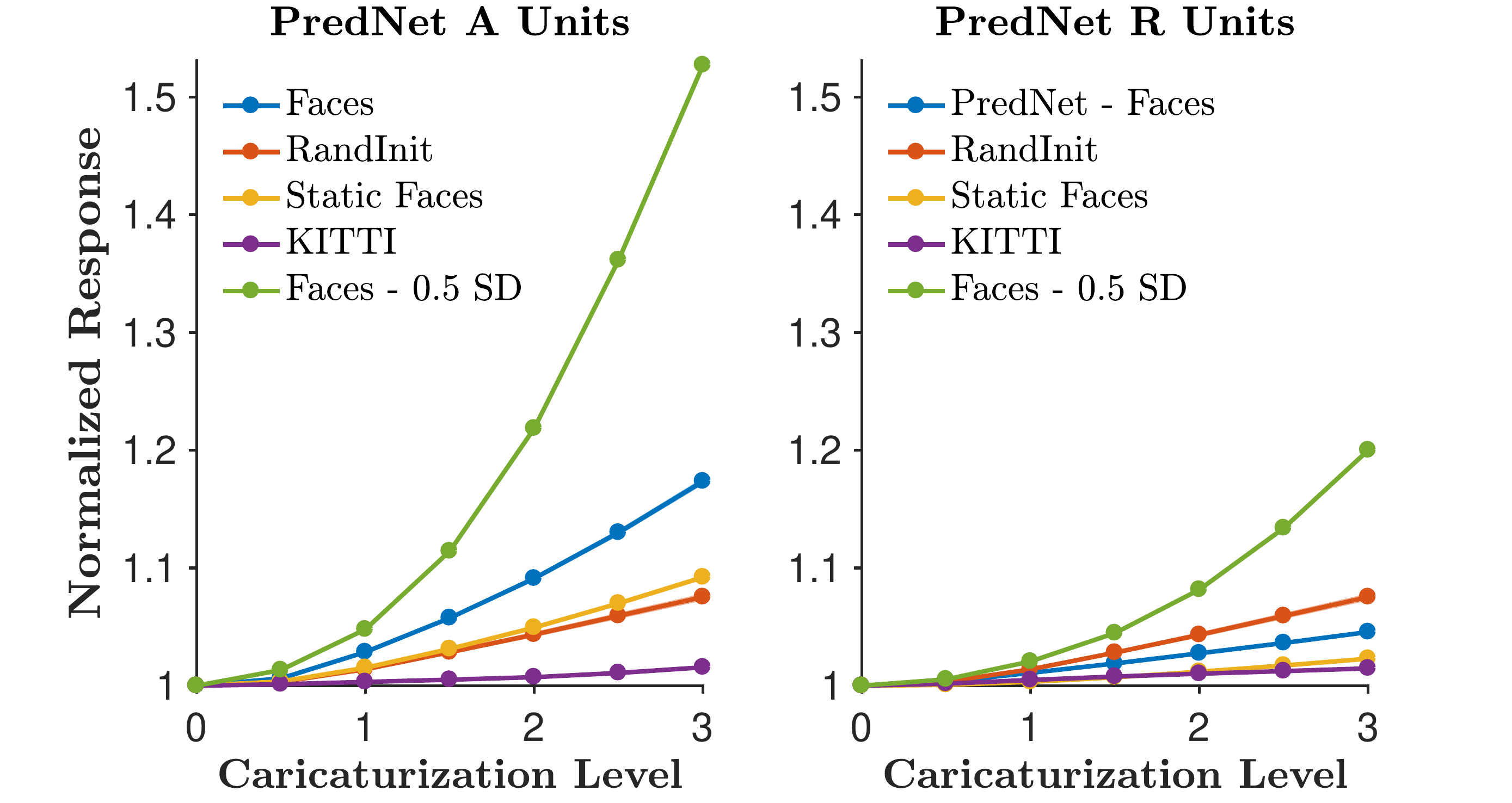}
	\end{center}
	\caption{Responses of PredNet $A$ and $R$ units to varying levels of caricaturized faces, trained in different settings. Faces - Rotating synthetic faces. RandInit - Random initial weights. Static Faces - Same collection of images used in Rotating Faces except presented statically. Faces 0.5 SD - Rotating faces except each face generated from a distribution with half of the original standard deviation. }
	\label{norm_faces_AR}
\end{figure}

For the faces generated for the norm-based coding experiment, a caricature level of, say $2$, corresponds to having all principal components with a of magnitude $2$ (either positive or negative).
The hyperparameters of the tested PredNet model were chosen to match those of the rotating faces model in the original paper \cite{Lotter_2017}. 
Fig. \ref{norm_faces_AR} shows the responses of the $A$ and $R$ units to the caricature faces.
Responses are calculated as an average per each layer, and then averaged across layers.
Training on rotating faces led to a much higher caricature response in the $A$ units, especially for training on faces generated with half of the original principal component standard deviation.
The lower standard deviation had a similar effect in the $R$ units, although training on the original rotating faces actually led to a smaller caricature response than the initial weights. 

The siamese VGG network used for the same/diff face identification task was constructed by taking the squared element-wise difference between the flattened features at the last convolutional layer for the two inputs, followed by a fully-connected, softmax classification layer. 

\subsection{Illusory Contours}

\begin{figure}[h]
	\begin{center}
		\includegraphics[width=1.0\textwidth]{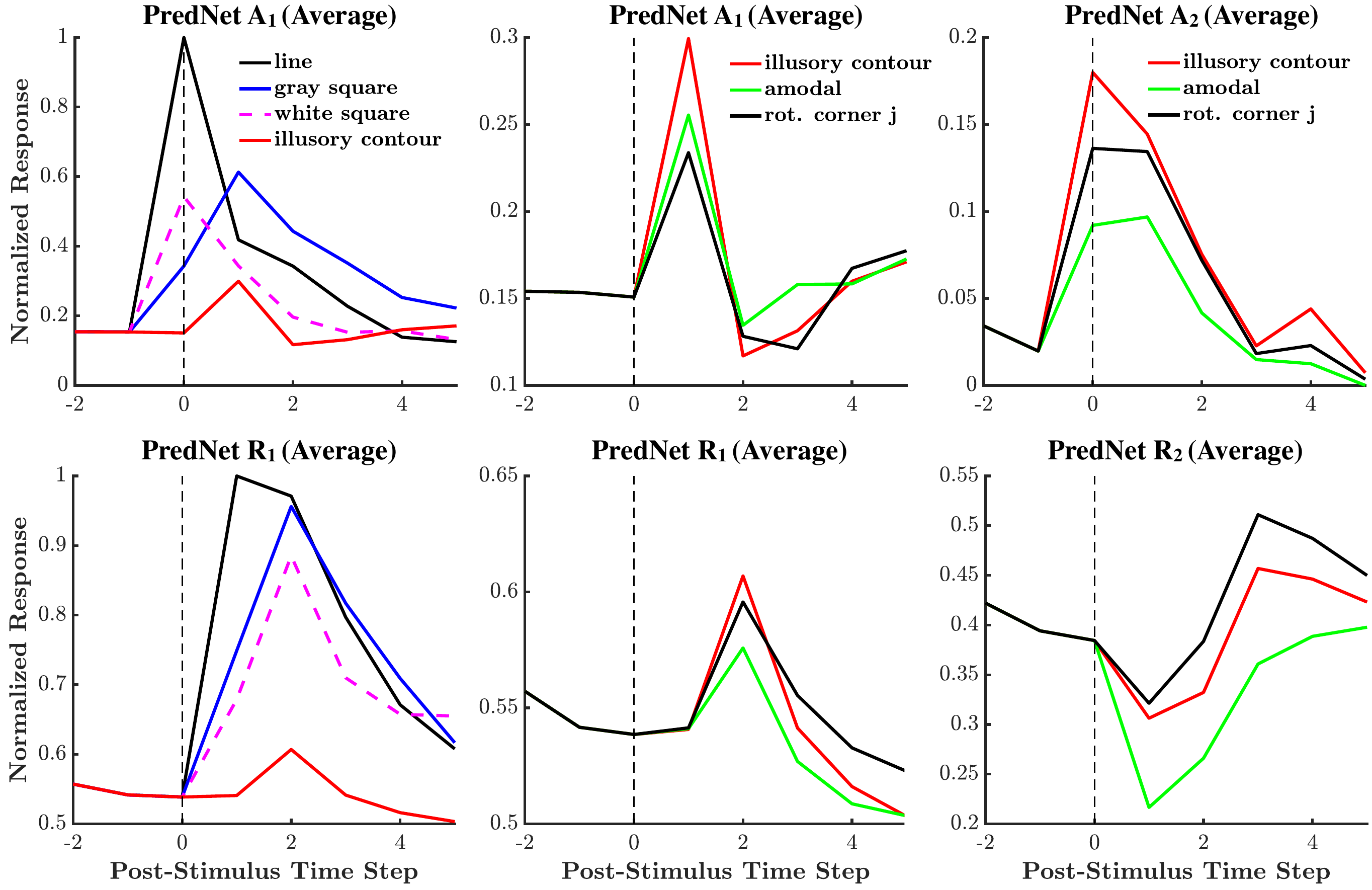}
	\end{center}
	\caption{Illusory contours responses for $A$ and $R$ units.}
	\label{illusory_contours_AR}
\end{figure}

Fig. \ref{illusory_contours_AR} contains the illusory contour response plots for the $A$ and $R$ layers.
The stimuli sequences consisted of $10$ time steps of the ``four circles'' image (see main text) followed by a test image for $10$ time steps.
The response to the illusory stimuli begins one time step after the response to the line square for all unit types in the first layer.

To quantify illusory responsiveness, we follow Lee et al. \cite{Lee_2001} in calculating the following two measures: $IC_a = \frac{R_i - R_a}{R_i + R_a}$ and $IC_r = \frac{R_i - R_r}{R_i + R_r}$, where $R_i$ is the response to the illusory contour (sum over stimulus duration), $R_a$ is the response to amodal stimuli, and $R_r$ is the response to the rotated J image.
These indices were calculated separately for each unit with a non-uniform response.
For both measures and all examined layers, the average across the layer was positive
 (significant in half of the calculations (Table~\ref{illusory_table})).

\begin{table}
  \caption{Illusory responsiveness measures for the units in Lee et al. \cite{Lee_2001} and the PredNet. $IC_a$ and $IC_r$ compare the response of the illusion to the amodal and rotated stimuli, respectively. Positive measures indicate a preference to the illusion. *$p<0.05$ (T-test)}
  \label{illusory_table}
  \centering
  \begin{tabular}{cccc|ccc|ccc|ccc}
    \toprule
    Source     & Layer  & $IC_A$   & $IC_R$  & Layer  & $IC_A$   & $IC_R$ & Layer  & $IC_A$   & $IC_R$ & Layer  & $IC_A$   & $IC_R$ \\
    \midrule
    Monkey A   & V1S   & $0.19$ & $0.31$  & V2S & $0.21$ & $0.11$ & V1D & $0.09$ & $0.11$ & V2D & $0.28$ & $0.24$   \\
    Monkey B   & V1S   & $0.10$    & $0.16$  & V2S & $0.08$ & $0.12$ & V1D & $0.04$ & $0.13$ & V2D & $0.07$ & $0.20$ \\
    PredNet   & $E_1$   & $0.09$  & $0.14$* & $E_2$   & $0.15$*  & $0.09$ & $R_1$ & $0.11$* & $0.04$ & $R_2$ & $0.12$* & $0.03$ \\
    PredNet   & $A_1$   & $0.03$  & $0.10$* & $A_2$   & $0.15$*  & $0.09$ \\
    \bottomrule
  \end{tabular}
\end{table}

\subsection{Flash-Lag Effect}

\begin{figure}[h]
	\begin{center}
		\includegraphics[width=1.0\textwidth]{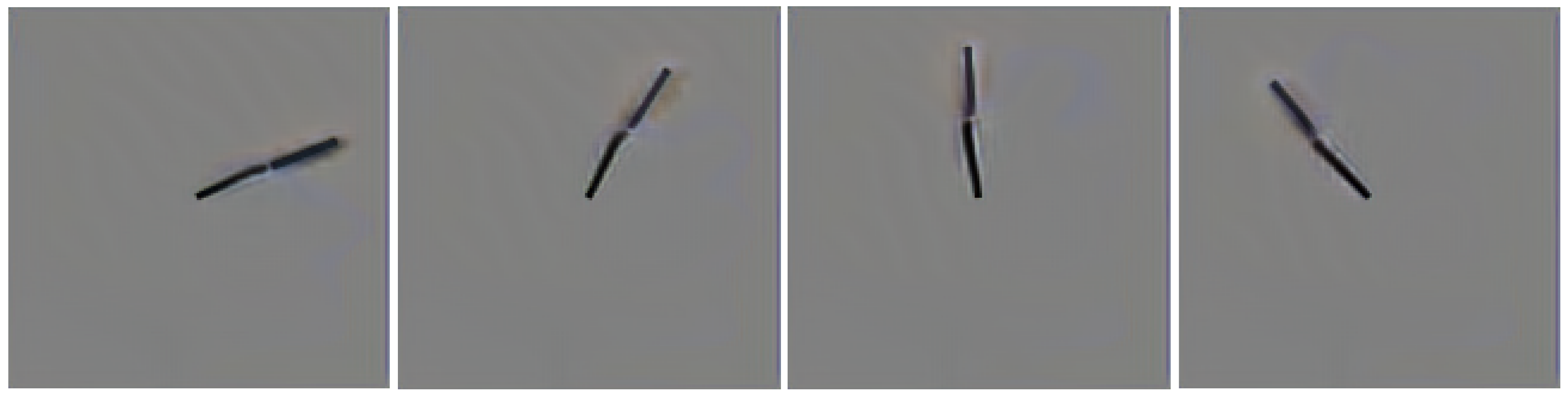}
	\end{center}
	\caption{Four consecutive post-flash predictions by the PredNet model following training on the KITTI dataset.}
	\label{flash_lag_post_flashes}
\end{figure}

The flash-lag stimulus was created with a rotation speed of $6$\degree per time step, with a flash every $6$ time steps for $3$ full rotations.
Angles of the predictions were quantified over the last two rotations.
The angles of the predicted bars were estimated by calculating the mean-squared error (MSE) between the prediction and a probe bar generated at $0.1$\degree increments and a range of centers, and taking the angle with the minimum mean-squared error.
Fig. \ref{flash_lag_post_flashes} contains additional predictions by the model after four consecutive flashes.

\end{document}